\documentstyle[preprint,aps,bezier]{revtex}
\begin{document}
\draft
\preprint{\parbox{4cm}{
\baselineskip=12pt
TMUP-HEL-9712\\ 
November, 1997\\
\hspace*{1cm}
}}
\title{A Composite Model of Quarks \\
        with the `Effective Supersymmetry' }
\author{Nobuchika Okada
\thanks{E-mail: n-okada@phys.metro-u.ac.jp}
\thanks{JSPS research fellow}}
\address{Department of Physics, Tokyo Metropolitan University\\
         Hachioji, Tokyo 192-0364, Japan}
\maketitle
\vskip 2.5cm
\begin{center}
{\large Abstract}
\vskip 0.5cm
\begin{minipage}[t]{14cm}
\baselineskip=19pt
\hskip4mm
We present a composite model of quarks with `effective supersymmetry'. 
The model is based on the gauge group 
 $(SU(2)_S \times SU(2)_M) \times (SU(2)_U \times SU(2)_C \times SU(2)_T)
 \times SU(5)_{SM}$, 
 where $SU(5)_{SM}$ is the standard model gauge group. 
In the dynamical supersymmetry breaking sector based on
 the gauge group $ SU(2)_S \times SU(2)_M $, 
 supersymmetry is dynamically broken.  
The preon sector is constructed by the model proposed by Nelson and Strassler. 
The fermion mass hierarchy among the up-type quarks originates from
 the $ SU(2)_U \times SU(2)_C \times SU(2)_T $ gauge dynamics. 
The supersymmetry breaking is mediated to 
 the minimal supersymmetric standard model sector 
 by the `preon' superfields which compose the quarks 
 in the first two generations. 
To obtain an experimentally acceptable mass spectrum, 
 the scalar quarks in the first two generations 
 need masses of order 10 TeV, while the other superpartners 
 need masses less than 1 TeV.
Therefore, the mass spectrum in our model is 
 one example of the `effective supersymmetry' model proposed by 
 Cohen, Kaplan and Nelson.  
\end{minipage}
\end{center}
\newpage
%
\section{Introduction}
\label{sec:int}

The standard model gives no understanding of 
 the origins of the electroweak symmetry breaking and 
 the fermion mass hierarchy. 
Furthermore, it is well known that there exists a naturalness problem 
 \cite{t'hooft} in the standard model. 
The electroweak scale cannot be stable with respect to quantum corrections 
 if new physics exists at very large scales.  

The naturalness problem is solved by minimal supersymmetric (SUSY) extension 
 of the standard model, the minimal supersymmetric standard model (MSSM). 
Although the extension gives no explanation of the origin of 
 the fermion mass hierarchy, 
 a new mechanism to break the electroweak symmetry is included. 
In this model, if SUSY is broken, 
 the electroweak symmetry breaking is caused 
 by the radiative correction due to the large top Yukawa coupling. 
This mechanism is known as the `radiative breaking scenario' \cite{rbs}. 
However, the supersymmetric mass term of the Higgs superfields,
 the $\mu$-term, should be roughly less than the electroweak scale, 
 so that this mechanism can work. 
Since the $\mu$-term is a free parameter, 
 its scale is not determined by the MSSM itself. 
This is known as the $\mu$ problem. 

To understand the MSSM more completely, 
 we should investigate three subjects as follows: 
(i) the mechanism of the SUSY breaking, 
(ii) a solution to the $\mu$ problem, 
(iii) an explanation of the large top Yukawa coupling 
 and the fermion mass hierarchy.  
  
The method proposed by Seiberg and co-workers \cite{seiberg} is noteworthy, 
 when we consider the dynamical scale generation and 
 the dynamical SUSY breaking in the $N=1$ SUSY gauge theory. 
By this method, we can evaluate the non-perturbative effect 
 of the strong gauge interaction and `exactly' determine 
 the dynamically generated superpotential. 
For example, there are many models [3-6] in which 
 the SUSY is broken by the strong gauge dynamics. 

On the other hand, the SUSY composite model 
 proposed by Nelson and Strassler \cite{ns} 
 is useful in understanding the large top Yukawa coupling 
 and the fermion mass hierarchy among the up-type quarks.  
They introduce the $SU(2)$ gauge interaction 
 with six doublet superfields, the `preon' superfields. 
The MSSM gauge group 
 $SU(5)_{SM} \supset SU(3)_c \times SU(2)_L \times U(1)_Y$ 
 is embedded in the $SU(6)$ global symmetry. 
Below the dynamical scale of the $SU(2)$ gauge interaction, 
 the preon superfields are confined into 
 the superfields of $\bf{5}$ and $\bf{10}$ representations 
 of $SU(5)_{SM}$, which are identified with the up-type Higgs 
 and the ten dimensional superfields in the conventional 
 SUSY $SU(5)$ GUT framework. 
The top Yukawa coupling is dynamically generated 
 by the non-perturbative effect of the $SU(2)$ gauge interaction,  
 and is naturally expected to be ${\cal{O}}(1)$. 
In addition, the $\mu$-term is also generated by strong dynamics. 
Unless the composite scale is much larger than the electroweak scale, 
 the $\mu$-term naturally becomes the electroweak scale. 
In the three generation case, three $SU(2)$ gauge interactions 
 are introduced at each generation. 
Then, the Yukawa couplings of the up-type quarks are generated by 
 each type of $SU(2)$ gauge dynamics. 
The small Yukawa couplings of the up and charm quarks 
 are effectively generated 
 through the mixing of the up-type Higgs superfields. 
The mass hierarchy among the up-type quarks can be understood 
 as the hierarchy among the three composite scales. 

An attempt to combine the SUSY composite model 
 with the dynamical SUSY breaking model 
 has been made in Ref.\cite{okada}. 
In this model, SUSY is dynamically broken, 
 the large top Yukawa coupling is dynamically generated, 
 and the radiative breaking scenario is realized. 
In addition, a new mechanism to give soft SUSY-breaking masses 
 to the superpartners in the MSSM was proposed. 
Here, the preon superfield plays the role of 
 the `messenger fields' \cite{dine}. 

However, the model cannot satisfy the naturalness criterion \cite{t'hooft}, 
 which is the motivation of the SUSY extension. 
In the model, the scalar top quark should have mass of order 10 TeV 
 to give an experimentally acceptable mass spectrum in the MSSM. 
Then, fine tuning is necessary to obtain the correct vacuum expectation 
 values of the Higgs bosons. 
In addition, since the composite scale is of order $10^{9}$ GeV, 
 an extremely small coupling constant is needed 
 to obtain the $\mu$-term of the electroweak scale.  

In this paper, we present a composite model of quarks 
 with `effective supersymmetry'. 
The model is based on the gauge group 
 $(SU(2)_S \times SU(2)_M) \times (SU(2)_U \times SU(2)_C \times SU(2)_T) 
 \times SU(5)_{SM}$. 
The dynamical SUSY breaking sector 
 is based on the gauge group $SU(2)_S \times SU(2)_M$.  
The strong $SU(2)_S \times SU(2)_M$ interaction dynamically breaks 
 SUSY and generates the $U(1)_R$ symmetry-breaking mass term 
 in the superpotential.  
The preon sector is constructed by the model
 proposed by Nelson and Strassler, 
 and is based on the gauge group $SU(2)_U \times SU(2)_C \times SU(2)_T$.  
The quarks in the $i$-th generation are composed 
 by the preon superfields which are confined by the strong $SU(2)_i$ 
 gauge interaction, 
 where $i=U,C$ and $T$ correspond to the first, second and third generation, 
 respectively. 
Since the gauge group $SU(2)_U \times SU(2)_C$ is embedded 
 in the global symmetry of the dynamical SUSY breaking sector, 
 the preon superfields in the first two generations 
 play the role of the messenger fields. 
To obtain an experimentally acceptable mass spectrum in the MSSM, 
 the scalar quarks in the first two generations 
 need masses of order 10 TeV. 
On the other hand, the other superpartners need masses less than 1 TeV. 
Therefore, the mass spectrum in our model is one example 
 of `effective supersymmetry' model 
 proposed by Cohen, Kaplan and Nelson \cite{ckn}. 

In section \ref{sec:cmp}, 
 we construct the preon sector 
 which is based on the model proposed by Nelson and Strassler. 
In section \ref{sec:dsb},
 the dynamical SUSY breaking sector is constructed. 
In section \ref{sec:feed}, 
 we discuss the mediation of the SUSY breaking to low energy, 
 and we estimate the mass spectrum in the MSSM. 
In section \ref{sec:ewb}, 
 we show that the electroweak symmetry breaking is realized 
 with an experimentally acceptable mass spectrum in the MSSM. 
In section \ref{sec:down}, 
 we consider the masses of the down-type quarks. 
We summarize our model in section \ref{sec:summary}.
%
%
\section{the preon sector}
\label{sec:cmp}

In this section we construct the preon sector, which is based on the model 
 proposed by Nelson and Strassler \cite{ns}. 
We introduce the $SU(2)_U \times SU(2)_C \times SU(2)_T$ gauge interaction, 
 which confines the preon superfields into the quark superfields 
 in three generations. 
Each type of dynamics based on the $SU(2)_i$ ($i=U,C,T$) gauge interaction 
 possesses the six doublet preon superfields, 
 where the subscripts $i=U,C$ and $T$ correspond
 to the first, second and third generation, respectively. 
The MSSM gauge group 
 is embedded in the $SU(6)$ global symmetry of the $SU(2)_i$ gauge dynamics. 

The particle contents are as follows:  
\begin{center}
\begin{tabular}{c|c|c|c}
 \hspace{1cm}& $~SU(2)_i$~ & $~SU(5)_{SM}~$ & $~U(1)_R~$   \\
\hline
$ P_i $      &  \bf{2} &  \bf{5}        & 1/4  \\
$ N_i $      &  \bf{2} &  \bf{1}        & 3/4  \\
\hline
$\bar{H}_i$    &  \bf{1} & $\bf{\bar{5}}$  & 1   \\
$\bar{\Phi}_i$  &  \bf{1} & $\bf{\bar{5}}$ & 1/2 \\
\end{tabular}
\end{center}
The superfield $\bar{H}_i$ is the down-type Higgs superfield, 
 and $\bar{\Phi}_i$ is the matter superfield including 
 the $SU(2)_L$ singlet quark of the down-type and the lepton doublet. 
In this paper, we use the $SU(5)$ GUT notation for simplicity. 
It is easy to decompose the notation into that for the MSSM 
 gauge group. 
In the following discussion, we consider only the quark sector in the MSSM. 

At high energy, where the $SU(2)_U \times SU(2)_C \times SU(2)_T$ 
 gauge interaction becomes weak, 
 most general tree-level superpotential is given by 
\begin{eqnarray}
  W_{\mbox{tree}} = \bar{H}_U \sum_{i=U,C,T} \alpha_i \left[P_i N_i \right] 
   + \bar{H}_C \sum_{i=C,T} \beta_i \left[P_i N_i \right]
   + \gamma \bar{H}_T \left[P_T N_T \right]  \; \; , 
 \label{nstree} 
\end{eqnarray}
where $\left[ \; \right]$ denotes the contraction of $SU(2)_i$ indices 
 by the $\epsilon$-tensor, and $\alpha_i$, $\beta_i$ and $\gamma$ are 
 dimensionless coupling constants. 
We redefine the flavor index of ${\bar H}_i$ as above 
 by the $SU(3)$ flavor rotation. 

At low energy, where the gauge interaction becomes strong, 
 the dynamical superpotential is generated as   
\begin{eqnarray}
  W_{\mbox{dyn}} = 
\sum_{i=U,C,T} \frac{1}{\Lambda_i^3} \mbox{Pf}( V_i) \; \; , 
 \label{nseff}
\end{eqnarray}
where $\Lambda_i$ is the dynamical scale of the $SU(2)_i$ gauge interaction, 
 and $V_i$ is a $6 \times 6$ antisymmetric tensor defined by 
\begin{eqnarray}
   V_i =  \left[ 
   \begin{array}{cc}
    \left[P_i P_i \right] & \left[P_i N_i \right]  \\
    \left[N_i P_i \right] & 0  \\
   \end{array}   \right]  
  \sim  \Lambda_i \left[
   \begin{array}{cc}
     {\bf 10}_i    &  H_i \\
     -H_i         &  0  \\
   \end{array}     \right]  \; \; . 
\end{eqnarray}
The fields ${\bf 10}_i$ and $H_i$ are the effective fields as follows: 
\begin{center}
\begin{tabular}{cclcc}
\hspace{1cm} &  \hspace{1cm}&  \hspace{1cm} & $ ~SU(5)_{SM}~ $ & \\
$ {\bf{10}}_i $ & $\sim $ & $\left[P_i P_i \right]/\Lambda_i $ & \bf{10}& \\
$ H_i $         & $\sim $ & $\left[P_i N_i \right]/\Lambda_i$ &\bf{5}& .      
\end{tabular}
\end{center}
We can identify ${\bf 10}_i$ and $ H_i $ with the superfield of 
 the {\bf 10} representation and the up-type Higgs superfield of 
 the {\bf 5} representation in the GUT framework, respectively.   
Then, we obtain the effective superpotential 
\footnote{ 
Since the first term in Eq. (\ref{nseff1}) can cause 
 too rapid proton decay, 
 we have to identify the $SU(2)_L$ singlet fields in ${\bf 10}_i$ 
 with new heavy fields to avoid this decay. 
The fields can have masses of order $\Lambda_i$, as considered by 
 Nelson and Strassler. 
In the following discussion,
 we consider only the quark sector, as already mentioned. 
}
\begin{eqnarray}
 W_{\mbox{eff}} =  \sum_{i=U,C,T} H_i {\bf{10}}_i {\bf{10}}_i 
   + \sum_{i=U,C,T} \alpha_i \Lambda_i \bar{H}_U H_i 
   +\sum_{i=C,T} \beta_i \Lambda_i \bar{H}_C H_i
   +\gamma \Lambda_T \bar{H}_T H_T  \; \; . 
 \label{nseff1} 
\end{eqnarray}
Note that the first term is nothing but the Yukawa coupling, 
 which is naturally ${\cal O}(1)$, and 
 the other terms are the $\mu$-terms of the Higgs superfields. 
By the strong $SU(2)_U \times SU(2)_C \times SU(2)_T$ dynamics, 
 the large Yukawa couplings and the $\mu$-terms are generated.  

Let us assume that the pair of up- and down-type Higgs superfields 
 ${\bar H}_T$ and $H_T$ is the lightest of all the Higgs superfields. 
Then, integrating out the heavy superfields from Eq. (\ref{nseff1}),  
 we obtain 
\begin{eqnarray} 
  W_{\mbox{eff}}= 
  \alpha  \frac{\Lambda_T}{\Lambda_U} H_T {\bf{10}}_U {\bf{10}}_U 
  + \beta \frac{\Lambda_T}{\Lambda_C}H_T {\bf{10}}_C {\bf{10}}_C  
  +  H_T {\bf{10}}_T {\bf{10}}_T 
  +  \gamma \Lambda_T \bar{H}_T H_T  \; \; , 
\label{mut}  
\end{eqnarray}
where $\alpha=(\alpha_C \beta_T)/(\alpha_U \beta_C) -\alpha_T/\alpha_U$
 and $\beta=-\beta_T/ \beta_C$. 
For simplicity, we take $\alpha_i \sim \beta_j$.  
Then $\alpha \sim \beta \sim {\cal{O}}(1)$. 
If $\langle H_T \rangle \neq 0$ is realized, 
 the masses of the up-type quarks are generated, 
 and we can obtain the mass relation
\begin{eqnarray}
 m_u : m_c :m_t \sim  
 \frac{\Lambda_T}{\Lambda_U} : \frac{\Lambda_T}{\Lambda_C} : 1 
 \; \; . 
\end{eqnarray}
Therefore, the mass hierarchy among the up-type quarks 
 originates from the hierarchy among the three composite scales, 
 $\Lambda_T \ll \Lambda_C \ll \Lambda_U$. 
Note that if $\Lambda_T$ is roughly the electroweak scale, 
 we obtain the $\mu$-term of the electroweak scale in Eq. (\ref{mut}). 
%
%
\section{the Dynamical SUSY Breaking sector}
\label{sec:dsb}

In order for the SUSY composite model introduced in the previous section 
 to be realistic, SUSY should be broken. 
We construct the dynamical SUSY breaking sector in this section.

It is pointed out that the theory of 
 the $SU(2)$ gauge interaction with four doublet and six singlet  
 superfields can cause the dynamical SUSY breaking \cite{iy}.  
The dynamical SUSY breaking sector in our model 
 has the extended structure of the theory. 
This sector is based on the $SU(2)_S \times SU(2)_M$ gauge group. 
To make our discussion clear, let us first 
 consider only the $SU(2)_S$ dynamics. 
The particle contents are as follows:
\begin{center}
\begin{tabular}{c|c|cc|c}
 \hspace{1cm} & $~SU(2)_S~$ & $~SU(2)_U~$ & $~SU(2)_C~$  & $~U(1)_R~$   \\
\hline 
$ Q_1 $  & \bf{2}      & \bf{2}  & \bf{1} &  0  \\
$ Q_2 $  & \bf{2}      & \bf{1}  & \bf{2} &  0  \\
\hline
$ Z  $        & \bf{1} & \bf{1} & \bf{1}  &  2  \\
$ Z^{\prime}$ & \bf{1} & \bf{1} & \bf{1}  &  2  \\
$ Z_5       $ & \bf{1} & \bf{1} & \bf{1}  &  2  \\
$ Z_D $       & \bf{1} & \bf{2} & \bf{2}  &  2  \\
$ N  $        & \bf{1}  & \bf{1} &  \bf{1} &  0
\end{tabular}
\end{center}
The gauge group $SU(2)_U \times SU(2)_C$ is embedded 
 in the global $SU(4)$ flavor symmetry of the $SU(2)_S$ dynamics.  
It is known that this symmetry breaks down to 
 the global $Sp(4)$ symmetry at low energy \cite{seiberg} 
 if the $SU(2)_U \times SU(2)_C$ gauge interaction is switched off. 
Then, we require the tree-level superpotential 
 to have the global $Sp(4)$ symmetry 
 when the gauge interaction is ignored. 
We define $Z$, $Z^\prime$ and $N$ as the singlet superfields, 
 and $Z_5$ and $Z_D$ as the {\bf 5} representation of this global symmetry. 

Without loss of generality, 
 tree-level superpotential is given by 
\begin{eqnarray}
   W_{\mbox{tree}}= & \lambda_Z & 
 Z ( \left[Q_1 Q_1\right] + \left[Q_2 Q_2 \right])/2 
  + \lambda_{Z^\prime}  Z^\prime 
  \left\{ ( \left[Q_1 Q_1\right] + \left[Q_2 Q_2 \right])/2 - \lambda_N N^2 
 \right\}  \nonumber \\
   &+& \lambda_5 \left\{ Z_5 
 (\left[Q_1 Q_1\right] - \left[Q_2 Q_2 \right])/2 
 + Z_D \left[Q_1 Q_2 \right]/2  \right\}  \; \; ,
 \label{dsbtree1} 
\end{eqnarray}
where $\left[ \; \right]$ denotes the contraction of the $SU(2)_S$ indices. 
The representation of $\left[ Q_i Q_j \right]$ ($i,j=1,2$) 
 is decomposed into {\bf 1}+{\bf 5} representation 
 of the $Sp(4)$ global symmetry, as above.  

At low energy, where the $SU(2)_S$ gauge interaction becomes strong, 
 the moduli space is dynamically deformed  
 in order to satisfy the condition Pf$\hat{V}=\Lambda_S^4$,  
 where $\hat{V}$ is $4 \times 4$ antisymmetric tensor given by 
\begin{eqnarray}
  \hat{V} =  \left[ 
   \begin{array}{cc}
    \left[Q_1 Q_1 \right]   &  \left[ Q_1 Q_2 \right] \\
    \left[Q_2 Q_1 \right]   &  \left[ Q_2 Q_2 \right]  \\
   \end{array}   \right]  
  \sim  \Lambda_S \left[
   \begin{array}{cc}
     V+V_5    &  V_D    \\
     -V_D     &  V-V_5  \\
   \end{array}     \right]  \; \; . 
\end{eqnarray}
Here, $ \Lambda_S $ is the dynamical scale 
 of the $SU(2)_S$ gauge interaction.
The fields $V$, $V_5$ and $V_D$ are the effective fields, as follows: 
\begin{center}
 \begin{tabular}{cclccc}
\hspace{1cm}&\hspace{1cm}&&$~~~SU(2)_U$&$ SU(2)_C~$ & \\
$ V+V_5 $ & $\sim $ & $ \left[Q_1 Q_1 \right]/\Lambda_S $ 
& \bf{1}&  \bf{1} & \\
$ V-V_5 $ & $\sim $ & $ \left[Q_2 Q_2\right] /\Lambda_S $ 
& \bf{1}&  \bf{1} & \\
$ V_D $   & $\sim $ & $ \left[Q_1 Q_2\right] /\Lambda_S $ 
& \bf{2}&  \bf{2} & .  
 \end{tabular}
\end{center}
Since the condition Pf$\hat{V}=\Lambda_S^4$ contradicts 
 the SUSY vacuum conditions 
 required by the tree-level superpotential of Eq. (\ref{dsbtree1}), 
 supersymmetry is dynamically broken \cite{iy}. 

To obtain the effective superpotential,
 we eliminate one of the effective fields 
 with the condition Pf$\hat{V}=\Lambda_S^4$.   
Using the effective fields, the condition is described by 
\begin{eqnarray}
 V^2 -V_5^2 - \frac{1}{2} V_D^2 = \Lambda_S^2 
 \; \; .
\end{eqnarray}
Considering small fluctuations of $V$ 
 around $\langle V \rangle = \Lambda_S$, we obtain 
\begin{eqnarray}
     V \sim \Lambda_S + \frac{1}{2 \Lambda_S} V_5^2 
   + \frac{1}{4 \Lambda_S} V_D^2 \; \; .
\end{eqnarray}
Eliminating $V$ from Eq. (\ref{dsbtree1}),
 the effective superpotential is given by
\begin{eqnarray}
   W_{\mbox{eff}} \sim &\lambda_Z&  Z 
   \left( \Lambda_S^2+ \frac{1}{2}V_5^2 + \frac{1}{4}V_D^2  \right) 
  + \lambda_{Z^\prime} Z^\prime \left( \Lambda_S^2 + \frac{1}{2}V_5^2 
  + \frac{1}{4} V_D^2 - \lambda_N N^2  \right)  \nonumber \\
   & + & \lambda_5 \Lambda_S 
  \left( Z_5 V_5 + \frac{1}{2} Z_D V_D \right)
   \; \; . \label{dsbeff1}
\end{eqnarray}
This effective superpotential is 
 one of the types of the O'Raifeartaigh model \cite{o'r}. 
For small $\lambda_Z$ compared with $\lambda_5$, 
 supersymmetry is broken by 
 $\langle F_Z \rangle = \lambda_Z \Lambda_S^2$, 
 where $ F_Z $ is the $F$-component of $Z$. 

Note that the scalar potential derived 
 from Eq. (\ref{dsbeff1}) has a `pseudo-flat' direction; 
 namely, the potential remains minimum along an arbitrary value 
 of $\langle Z \rangle$
\footnote{ We use the same notation for both the superfield itself 
 and the scalar component of the superfield.  
}. 
This pseudo-flat direction is lifted up by the quantum correction 
 for the scalar potential of $Z$.   
If we consider the Yukawa coupling in Eq. (\ref{dsbeff1}),  
 $\langle Z \rangle =0 $ is expected \cite{hqu}.  
Then, the vacuum is realized at 
 $\langle F_Z \rangle \neq 0$, $\langle$other $F$-components$\rangle =0$, 
 $\langle N \rangle = \Lambda_S / \sqrt{\lambda_N}$ 
 and $\langle$other scalar components$\rangle =0$. 
Note that there is no $U(1)_R$ symmetry breaking mass term 
 of $V_D$ in the effective superpotential, 
 because $\langle Z \rangle = \langle Z^\prime \rangle =0$.  

A strong $SU(2)_M$ gauge interaction is introduced 
 to generate the mass term. 
We also introduce two doublet superfields, $\bar{Q}$ and $Q$, of 
 the $SU(2)_M$ gauge group. 
In addition to the effective superpotential of Eq. (\ref{dsbeff1}), 
 let us consider the new tree-level superpotential 
\begin{eqnarray}
 W^{\prime}_{\mbox{tree}} = 
 \lambda_M N \; \left[ \bar{Q} Q  \right] \; \; ,
\label{tree2}
\end{eqnarray} 
where $\left[ \; \right]$ denotes the contraction of the $SU(2)_M$ indices 
 by the $\epsilon$-tensor. 
The vacuum is realized with the same vacuum expectation values 
 as discussed above 
 and enforcing the condition 
 $\langle \bar{Q} \rangle = \langle Q \rangle =0$. 

However, we should consider the non-perturbative effect of 
 the strong $SU(2)_M$ gauge interaction at low energy. 
Then, the effective superpotential is given by 
\begin{eqnarray}
 W^{\prime}_{\mbox{eff}}= \lambda_M \Lambda_M N M 
  + \frac{\Lambda_M^4}{M} \; \; ,  
\label{dsbeff2} 
\end{eqnarray}
where $\Lambda_M$ is the dynamical scale 
 of the $SU(2)_M$ gauge interaction, 
 and $M \sim \left[\bar{Q}Q \right] / \Lambda_M$ 
 represents the effective fields. 
Now we obtain the total effective superpotential 
 $\tilde{W}_{\mbox{eff}}= W_{\mbox{eff}}+W^{\prime}_{\mbox{eff}}$ 
 in the dynamical SUSY breaking sector. 

We can see that $\langle Z^{\prime} \rangle \neq 0$ is realized
 from the total superpotential. 
Indeed, from the two conditions  
 $\partial \tilde{W}_{\mbox{eff}}/ \partial M = 0$ 
 and $\partial \tilde{W}_{\mbox{eff}}/ \partial N = 0$, we obtain 
\begin{eqnarray}
 \langle M \rangle &=& \sqrt{\frac{1}{\lambda_M \langle N \rangle}}
 \Lambda_M^{3/2} 
 = \frac{\lambda_N^{1/4}}{\lambda_M^{1/2}}  
   \frac{\Lambda_M^{3/2}}{\Lambda_S^{1/2}} \; \; ,  \nonumber \\
 \langle Z^{\prime} \rangle &=& 
 \frac{\lambda_M}{2 \lambda_{Z^\prime} \lambda_N} 
 \frac{\langle M \rangle}{\langle N \rangle} \Lambda_M
 = \frac{\lambda_M^{1/2}}{2 \lambda_{Z^\prime}\lambda_N^{1/4}}
   \frac{\Lambda_M^{5/2}}{\Lambda_S^{3/2}}  \; \; . 
 \label{vac}
\end{eqnarray}
The $U(1)_R$ symmetry breaking mass term 
 for the superfield $V_D$ is generated. 

%
%
\section{The mediation of supersymmetry breaking and 
 the mass spectrum of the superpartners in the MSSM}
\label{sec:feed}

In this section, we investigate the mediation of the supersymmetry breaking. 
This breaking is mediated to the preon sector and the MSSM sector 
 by the $SU(2)_U \times SU(2)_C$ and $SU(5)_{SM}$ gauge interactions. 
Then, all of the superpartners in the MSSM obtain 
 soft SUSY-breaking masses. 

Let us extract the parts including $V_D$ 
 from the total effective superpotential $\tilde{W}_{\mbox{eff}}$ 
 given in the previous section. 
These are given by 
\begin{eqnarray}
W_{\mbox{mess}} = \frac{1}{4} (\lambda_Z Z + \lambda_{Z^\prime} Z^\prime)
 V_D^2  + \frac{1}{2} \lambda_5 \Lambda_S Z_D V_D\; \; . 
\label{mess1}
\end{eqnarray}
Because $\langle F_Z \rangle \neq 0 $ 
  and $ \langle Z^\prime \rangle \neq 0 $, 
  the superfield $V_D$ plays the role of the `messenger fields' \cite{dine} 
  to give soft SUSY-breaking masses to 
  the $SU(2)_U \times SU(2)_C$ charged particles. 
In the following discussion,
 we assume $\lambda_Z \langle F_Z \rangle \ll 
 (\lambda_{Z^\prime} \langle Z^\prime \rangle)^2$ 
 and  $\lambda_5 \Lambda_S \ll \lambda_{Z^\prime} 
 \langle Z^\prime \rangle$, for simplicity. 

Through the 1-loop radiative correction by the messenger fields $V_D$, 
 the $SU(2)_U$ and $SU(2)_C$ gauginos obtain the soft SUSY-breaking masses  
\begin{equation}
  m_{\lambda_i} \sim \frac{\alpha_{i}}{4 \pi} \frac{F}{m} \; \; , 
 \label{gauginomass}
\end{equation}
where $\alpha_{i}=g_i^2/4\pi$ is the coupling constant of 
 the $SU(2)_i$ gauge interaction ($i=U,C$), 
 $F=\lambda_Z \langle F_Z \rangle$ 
 and $m=\lambda_{Z^\prime} \langle Z^\prime \rangle$.  

The scalar preons, $\tilde{P}_i$ and $\tilde{N}_i$ ($i=U,C$), also 
 obtain masses through the 2-loop radiative correction,  
\begin{equation}
m_{\tilde{P}_i}^2=m_{\tilde{N}_i}^2 \sim \frac{3}{2} 
\left( \frac{\alpha_i}{4 \pi} \right)^2 \left( \frac{F}{m} \right)^2 
\label{spreonmass} \; .  
\end{equation}
The scalar preon masses are of the same order 
 as the gaugino masses. 
Since the scalar preons have masses, the composite scalars, 
 $\tilde{\bf 10}_i$ and the up-type Higgs boson $H_i$, also have 
 the soft SUSY-breaking masses. 
For simplicity, 
 we assume that the masses of the composite scalars are twice 
 as large as the corresponding scalar preon masses.

Furthermore, the gauginos in the MSSM obtain 
 the soft SUSY-breaking masses through 
 the radiative correction including the preon superfields $P_i$, 
 since the preon superfield $P_i$ has not only $SU(2)_i$ charge 
 but also $SU(5)_{SM}$ charge. 
By the 3-loop diagram in Fig. 1 \cite{kitazawa} \cite{randall}, 
 the masses are given by 
\begin{equation}
 m_{\lambda_N} \sim \frac{3 \alpha_N}{8 \pi} \frac{F}{m} 
 \sum_{i=U,C} 
  \left( \frac{\alpha_i}{4 \pi} \right)^2 
 \; \ln\left(\frac{m}{\Lambda_i}
  \right) \label{ewgaugino} \; \; ,
\end{equation}
where $\alpha_N$ ($N=1,2,3$) is the coupling constant 
 of the $SU(N)$ gauge interaction in the standard model  
\footnote{
In this calculation, an infrared divergence occurs.
This implies that the mass at zero momentum cannot be defined. 
We have taken $\Lambda_i$ as the cutoff parameter 
 of the infrared divergence,  
 since the preon superfields $P_i$ are confined 
 at an energy lower than $\Lambda_i$. 
}
. 
Note that the preon superfields also play the role of 
 the `messenger fields'. 

Using the above results, we can estimate the mass spectrum in the MSSM.  
However, note that the results cannot be reliable at low energy, 
 since the $SU(2)_i$ gauge interaction becomes strong there. 
For simplicity, we take $\alpha_H/4 \pi =1$ as a rough estimation. 
According to the above assumptions, we obtain 
\begin{equation}
 m_{\tilde{\bf 10}_i}=m_{H_i} \sim  2 m_{\tilde{P}_i} \sim 
  \sqrt{6} \left( \frac{F}{m} \right)\; \; . 
\label{squarkmass}
\end{equation}
In the following discussion, we regard 
 $\tilde{m} \equiv m_{\tilde{\bf{10}}_i}$ as a free parameter. 
Then, the MSSM gaugino masses are rewritten as
\begin{equation}
 m_{\lambda_N} \sim \frac{\sqrt{6}\alpha_N}{16 \pi} \; \tilde{m} \; 
 \sum_{i=U,C} \ln (m/\Lambda_i) 
 \sim \frac{3\alpha_N}{2} \; \tilde{m} \; \; ,
\label{smgauginomass}
\end{equation}
where we fix the value of the free parameter $m$ 
 such that $ \sum_{i=U,C} \ln (m/\Lambda_i) =4 \sqrt{6} \pi $. 

Since the MSSM gauginos and the scalar quarks in the first 
 two generations have soft SUSY-breaking masses, 
 the other scalar partners in the MSSM also obtain masses 
 through radiative corrections by the gauginos and the scalars. 
The scalar partner of ${\bf 10}_T$ obtains a mass
 mainly by the 1-loop correction due to the $SU(3)_c$ gaugino 
 and the 2-loop corrections 
 due to $\tilde{\bf{10}}_U$ and $\tilde{\bf{10}}_C$:   
\begin{equation}
 m_{\tilde{\bf{10}}_T}^2 \sim \frac{8 \alpha_3}{3 \pi} \; 
  m_{\lambda_3}^2 \; \ln (\Lambda_T/m_{\lambda_3}) 
  -  \frac{2 \alpha_3^2}{\pi^2}\; \tilde{m}^2 \; \ln (\Lambda_T/\tilde{m})
 \; \; . 
\label{stop}
\end{equation}
The Higgs bosons obtain masses
 mainly by the 1-loop correction due to the $SU(2)_L$ gaugino 
 and the 2-loop corrections 
 due to $\tilde{\bf{10}}_U$ and $\tilde{\bf{10}}_C$: 
\begin{equation}
 m_{\bar{H}_T}^2 \sim m_{H_T}^2 \sim \frac{3 \alpha_2}{2 \pi}\; 
 \; m_{\lambda_2}^2 \; \ln (\Lambda_T/m_{\lambda_2}) 
 -\frac{9 \alpha_2^2}{8 \pi^2}\; \tilde{m}^2 
  \; \ln (\Lambda_T/\tilde{m}) \; \; .
\label{higgs}
\end{equation}
Note that, since the 2-loop corrections give negative contributions 
 to the mass squared, our model potentially has a problem 
 of $SU(3)_c$ color breaking \cite{mura}.  
However, we can avoid this problem 
 by fixing the parameters such as $m \gg \Lambda_i$ 
 (see Eq. (\ref{smgauginomass})). 
The soft SUSY-breaking $B$-term, $B \bar{H}_T H_T + \mbox{h.c.}$,  
 in the Higgs potential is induced mainly by the $SU(2)_L$ gaugino loop, 
 and we obtain 
\begin{equation}
 B \sim \frac{3 \alpha_2}{2 \pi}\; m_{\lambda_2} \; \mu 
 \ln (\Lambda_T/m_{\lambda_2}) \; \; ,
\label{bterm}
\end{equation}
where $\mu = \gamma \Lambda_T$.  

Here, we list the masses of the superpartners 
 to be used in the following discussion: 
\begin{eqnarray}
m_{\lambda_3} &\sim & 1.8 \times 10^{-1}\; \tilde{m} \; \; ,\nonumber \\
m_{\lambda_2} &\sim & 5.0 \times 10^{-2} \; \tilde{m}\; \; , \nonumber \\
m_{{\tilde{\bf{10}}}_T} &\sim&  7.8 \times 10^{-2} \; \tilde{m} \; \; ,
 \nonumber \\
m_{\bar{H}_T} \sim m_{H_T} &\sim&  8.1 \times 10^{-3} \; \tilde{m} \; \; ,
\nonumber \\
B &\sim& 2.8 \times 10^{-3} \; \mu \; \tilde{m} \; \; .
\label{massvalue}
\end{eqnarray}
We have used $\alpha_3 \sim 0.12$ and $\alpha_2 \sim 0.033$ here, 
 and we take $\Lambda_T/m_{\lambda_3} \sim 10$, for simplicity. 

To obtain an experimentally acceptable mass spectrum in the MSSM \cite{pdg}, 
 it is required that the scalar quarks in the first two generations 
 be very heavy; namely,  $\tilde{m} \geq \cal{O}$(10 TeV). 
However, note that this value implies no contradiction 
 with the naturalness criterion \cite{t'hooft}. 
Since the Yukawa coupling constants of these quarks are very small, 
 the contribution to the self-energy of the Higgs bosons 
 remains less than the electroweak scale, 
 even if $\tilde{m} \geq \cal{O}$(10 TeV). 
On the other hand, the other superpartners have masses 
 less than 1 TeV. 
Therefore, the mass spectrum in our model is one example
 of the `effective supersymmetry' model proposed 
 by Cohen, Kaplan and Nelson \cite{ckn}. 
%
%
\section{The electroweak symmetry breaking }
\label{sec:ewb}

In order for our model to be realistic, 
 the electroweak symmetry should be broken. 
In this section, we investigate whether the symmetry breaking is 
 realized or not by using the mass spectrum obtained the previous section. 

The Higgs potential is given by 
\begin{eqnarray}
  V =& & (\mu^2+m^2_{\bar{H}_T}) |\bar{H_T}|^2 
   + (\mu^2+m^2_{H_T}) |H_T|^2 +( B \bar{H}_T H_T+ \mbox{h.c.}) \nonumber \\
  &+& \frac{1}{2} \left(\frac{M_Z}{v} \right)^2 
 (|\bar{H}_T|^2- |H_T|^2)^2 \; \; , 
 \label{ewtree}
\end{eqnarray}
where $M_Z=91$ GeV is the Z-boson mass, and $v=246$ GeV. 
Using the result from Eq. (\ref{massvalue}), 
 we can obtain 
\begin{eqnarray}
(\mu^2+m^2_{\bar{H}_T})(\mu^2+m^2_{H_T})-B^2 > 0 \nonumber
\end{eqnarray}
for arbitrary values of $\mu$ and $\tilde{m}$. 
Therefore, the electroweak symmetry is not broken 
 in this potential. 

Then, we consider the radiative correction of the Higgs potential 
 due to the large top Yukawa coupling. 
The 1-loop effective potential induced by the top and scalar top quarks 
 is given by 
\begin{equation} 
 V_{\mbox{1-loop}}= \frac{3}{8 \pi^2} \int^{\Lambda_T^2}_{0} dk_E^2 \; 
 k_E^2 \left[ \ln \left( 1+ \frac{g_t^2 |H_T|^2}{k_E^2 
 +m_{\tilde{\bf{10}}_T}^2} \right)
 -\ln \left( 1+ \frac{g_t^2 |H_T|^2}{k_E^2}  \right) \right] \; \; , 
\label{1loop}
\end{equation}
where we use $\Lambda_T$ as a cutoff parameter 
 because of the compositeness of the top quark superfield 
 and the up-type Higgs boson. 
In Eq. (\ref{1loop}), the first and second terms in the square brackets 
 are given by the loop corrections of the scalar top and top quarks, 
 respectively. 
One can check that $V_{\mbox{1-loop}} \rightarrow 0$ in the SUSY limit 
 $m_{\tilde{\bf 10}_T}^2 \rightarrow 0$. 
Now, we obtain the effective potential 
 $V_{\mbox{eff}}=V +V_{\mbox{1-loop}}$ 
 from Eq. (\ref{ewtree}) and (\ref{1loop}). 

In order for the electroweak symmetry to be broken, 
 two minimization conditions, 
 $\partial V_{\mbox{eff}}/ \partial \bar{H}_T=0$ 
 and $\partial V_{\mbox{eff}}/ \partial H_T =0$, should be satisfied 
 with nonzero values of $\langle \bar{H}_T\rangle$ 
 and $\langle H_T \rangle$.
In addition, these vacuum expectation values should satisfy the relation 
 $\langle \bar{H}_T \rangle^2 + \langle H_T \rangle^2 = v^2 / 2$. 
Then, the minimization conditions are described by 
\begin{eqnarray}
& & \mu^2 + m_{\bar{H}_T}^2 + \frac{1}{2} M_Z^2 \cos 2\beta  
+ B \tan \beta  = 0 \; \;  , 
 \label{cond1} \\
& &\mu^2  +m_{H_T}^2  
 - \frac{1}{2} M_Z^2 \cos 2\beta+B \cot \beta
  \nonumber \\
&+& \frac{3}{4 \pi^2} \frac{m_t^2}{v^2 \sin^2 \beta} \left[ 
  m_t^2 \ln \left(\frac{\Lambda_T^2 + m_t^2}{m_t^2}\right) 
- (m_{\tilde{\bf 10}_T}^2 +m_t^2) \ln \left 
 (\frac{\Lambda_T^2 + m_{\tilde{\bf{10}}_T}^2 +m_t^2}
 {m_{\tilde{\bf{10}}_T}^2 + m_t^2} \right)
\right]= 0  \; \; ,
 \label{cond2}
\end{eqnarray}
where we use the definitions 
 $\langle \bar{H}_T \rangle = v \cos \beta /\sqrt{2}$, 
 $\langle H_T \rangle = v \sin \beta /\sqrt{2}$ 
 and $m_t^2=g_t^2 \langle H_T \rangle^2$ for the top quark mass. 

We will show that there is a realistic solution 
 by substituting realistic values
 of the parameters into Eqs. (\ref{cond1}) and (\ref{cond2}). 
There are four independent unknown parameters, 
 $\mu$, $\tilde{m}$, $\tan \beta$ and $\Lambda_T$. 
When we fit the values of two of these parameters as
 $\mu = -210$GeV and $\tilde{m} = 6$ TeV, 
 Eqs. (\ref{cond1}) and (\ref{cond2}) are regarded as 
 equations to determine $\tan \beta$ and $\Lambda_T$, respectively. 
Then, we can find the solution 
\footnote{
Our assumption $\Lambda_T/m_{\lambda_3} \sim 10$ is
 satisfied by this solution.}
\begin{eqnarray}
\tan \beta \sim 12 \; , \; \; 
\Lambda_T \sim 11 \; \mbox{TeV} \; \; ,  \nonumber 
\end{eqnarray}
where we take $m_t=175$ GeV. 
The values of $|\tan \beta|$ and $\Lambda_T$ become 
 large (small) as $|\mu |$ ($\tilde{m}$) becomes large. 

The input value $\tilde{m}=6$ TeV gives an experimentally acceptable
 mass spectrum in the MSSM; 
 for example, $m_{\lambda_3}\sim 1$ TeV from Eq. (\ref{massvalue}).  
From this solution, we can obtain the hierarchy  
 $m_{\tilde{\bf 10}_T} \ll \tilde{m} \sim \Lambda_T 
 \ll \Lambda_C \ll \Lambda_U$, 
 which should be satisfied for the mechanism 
 of Nelson and Strassler to work. 
This hierarchy also forbids the Higgs bosons 
 (except for ${\bar H}_T$ and $H_T$) 
 to obtain nonzero vacuum expectation values  
 because of their very large $\mu$-terms.  
On the other hand, the $\mu$-term of ${\bar H}_T$ and $H_T$ 
 can become of the electroweak scale 
 with a not so small parameter $\gamma \sim 10^{-2}$. 
As a result, the radiative-breaking scenario is realized 
 with a realistic solution. 
%
%
\section{The masses of the down-type quarks}
\label{sec:down}

In the previous sections, the masses of the down-type quarks were
 not considered. 
In this section, we discuss the method to generate the masses 
 of the down-type quarks. 

Since the quark doublets are composed of two preons, 
 we cannot introduce the Yukawa coupling of the down-type quarks 
 with renormalizable terms in the superpotential. 
Following to Nelson and Strassler \cite{ns}, 
 let us introduce heavy fields $\bar{S}_i$ and $S_i$ 
 as follows: 
\begin{center}
\begin{tabular}{c|cc}
 \hspace{1cm}    & $~SU(2)_i~$ &  $~U(1)_R~$   \\
\hline 
$ \bar{S_i} $   & \bf{2}      & 3/4  \\
$ S_i  $          & \bf{2}      & 5/4   \\
\end{tabular}
\end{center}
Here, $i=U,C$ and $T$ correspond to each generation. 
For simplicity, we discuss only the third generation. 

In addition to the tree-level superpotential of Eq. (\ref{nstree}), 
 we can introduce the superpotential
\begin{eqnarray}
  W=\lambda_{\bar{S}_T}\bar{H}_T \left[P_T \bar{S}_T \right] 
 + \lambda_{S_T} \left[P_T S_T \right] \Phi_T 
 - m_{\bar{s} s} \bar{S}_T S_T \; \; , 
\label{intermeditate}
\end{eqnarray}
where $\lambda_{{\bar S}_T}$ and $ \lambda_{S_T}$ are 
 dimensionless coupling constants, and 
 $m_{\bar{s} s}$ is the mass parameter.
Integrating out the heavy fields, we obtain 
\begin{eqnarray}
 W= \frac{\lambda_{\bar{S}_T}  \lambda_{S_T}}{m_{\bar{s} s}} 
    \bar{H}_T  \left[P_T P_T \right] \Phi_T \; \; . 
\end{eqnarray}
Then, after the confinement of $ \left[P_T P_T \right]$, 
 this superpotential gives the Yukawa coupling of the bottom quark, 
\begin{eqnarray}
 W= \lambda_{{\bar S}_T}  \lambda_{S_T}  
 \frac{\Lambda_T}{ m_{\bar{s} s}} 
 \bar{H}_T \Phi_T {\bf{10}}_T  \; \; . 
\end{eqnarray}
Note that the mass hierarchy between the top and bottom quarks is 
 obtained by $\Lambda_T \ll  m_{\bar{s} s}$. 
This mechanism was introduced by Nelson and Strassler. 

However, we should explain the origin of the mass parameter 
 $m_{\bar{s} s}$. 
In addition to the effective superpotential $\tilde{W}_{\mbox{eff}}$ 
 in section {\ref{sec:dsb}, let us introduce the superpotential 
\begin{eqnarray}
W= \lambda_{\bar{s} s} N \left[ {\bar S}_T S_T \right] \; \; .
\end{eqnarray}
Note that the vacuum discussed in section {\ref{sec:dsb} 
 is not changed by this introduction. 
We can obtain the mass term of ${\bar S}_T$ and $S_T$ 
 by $\langle N \rangle = \Lambda_S / \sqrt{\lambda_N}$. 
The origin of the mass term is the dynamical scale 
 of the $SU(2)_S$ gauge interaction.  
Therefore, the mass hierarchy between the top and bottom quarks 
 originates from the hierarchy between the composite 
 and the dynamical SUSY breaking scales.

It is easy to extend our discussion to the three generation case. 
However, our model has no mechanism to explain the mass hierarchy 
 between the down-type quarks in the different generations. 
The explanation of the masses of the down-type quarks remains 
 an open problem in our model. 
%
%
\section{Summary}
\label{sec:summary}
 
We have constructed a composite model of quarks  
 with `effective supersymmetry'. 
The model is based on the gauge group 
  $(SU(2)_S \times SU(2)_M) \times (SU(2)_U \times SU(2)_C \times SU(2)_T)
 \times SU(5)_{SM}$.  

In the dynamical SUSY breaking sector, 
 supersymmetry is dynamically broken by the strong $SU(2)_S$ gauge dynamics. 
This dynamics also generates the $SU(2)_U \times SU(2)_C$ bi-doublet 
 superfield $V_D$ as a composite field. 
The $U(1)_R$ symmetry-breaking mass of $V_D$ is generated by the 
 strong $SU(2)_M$ gauge dynamics. 
Then, the bi-doublet superfield plays the role of the messenger field 
 to mediate the SUSY breaking to the preon sector 
 in the first two generations. 

In the preon sector, the preon superfields 
 are confined in the up-type Higgs superfield $H_i$ 
 of the $\bf{5}$ representation and the superfield ${\bf 10}_i$
 of the $\bf{10}$ representation by the non-perturbative effect 
 of the $SU(2)_i$ gauge interaction. 
A large top Yukawa coupling is dynamically generated. 
On the other hand, small Yukawa couplings of the up and charm 
 quarks are effectively generated 
 through the mixing of the up-type Higgs superfields. 
The mass hierarchy among the up-type quarks 
 originates from the hierarchy among the three composite scales. 

SUSY breaking is mediated to the preon sector 
 and the MSSM sector through the radiative correction 
 by the $SU(2)_U \times SU(2)_C$ bi-doublet superfield $V_D$ and 
 the preon superfield $P_i\; (i=U,C)$.  
In particular, the MSSM gauginos obtain masses 
 in a manner described by the 3-loop diagram in Fig. 1. 
Here, it is crucial for the preon superfield $P_i$ to have charges 
 both of the $SU(2)_i$ and the MSSM gauge groups, 
 and it plays the role of the `messenger fields'. 
In order to obtain an experimentally acceptable mass spectrum 
 in the MSSM, the scalar quarks in the first two generations 
 need masses of order 10 TeV.  
On the other hand, the other superpartners need masses 
 less than 1 TeV. 
Therefore, the mass spectrum in our model is an example
 of the `effective supersymmetry' model 
 of Cohen, Kaplan and Nelson.  
 
In our model, the electroweak symmetry is broken 
 by the radiative breaking scenario. 
The conditions to realize this scenario 
 are satisfied with an experimentally acceptable mass spectrum in the MSSM. 
Our result is consistent with the condition 
 for the mechanism of Nelson and Strassler to work.  
The $\mu$-term of ${\bar H}_T$ and $H_T$ 
 becomes the electroweak scale without a very small parameter, 
 because of the solution $\Lambda_T \sim 10$ TeV.  

The masses of the down-type quarks are introduced 
 by the higher dimensional terms as proposed by Nelson and Strassler. 
In our model, the origin of the mass term of ${\bar S}_T$ and $S_T$ 
 is the dynamical scale of the $SU(2)_S$ gauge interaction. 
Then, the mass hierarchy between the top and bottom quarks  
 originates from the hierarchy between the composite 
 and the dynamical SUSY breaking scales. 
However, there is no mechanism to explain the hierarchy 
 among the masses of the down-type quarks 
 in different generations. 
\acknowledgments
The author would like to thank Noriaki Kitazawa for useful comments. 
This work was supported in part by a Grant-in-Aid 
 for Scientific Research from the Ministry of Education, 
 Science and Culture and a Research Fellowship of the Japan Society 
for the Promotion of Science for Young Scientists. 
%

%

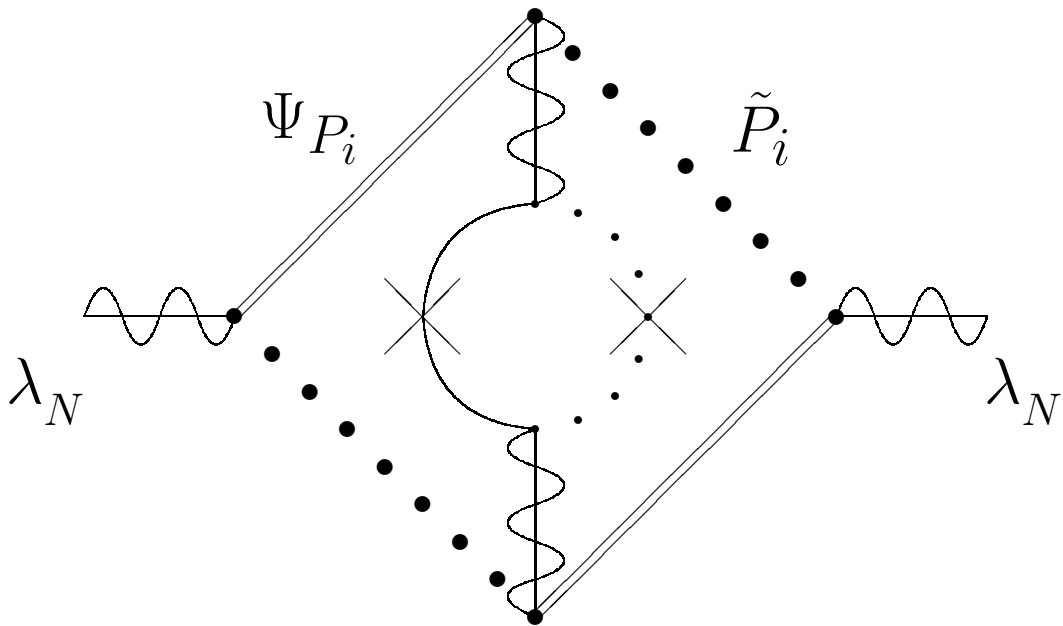
\begin{figure}
\caption{The 3-loop diagram for the soft SUSY-breaking masses 
 of the MSSM gauginos. 
 The double solid and large-dotted lines denote the propagators
 of the fermion $\Psi_{P_i}$ and scalar $\tilde{P_i}$ components 
 of the preon superfield $P_i$, respectively. 
 The solid and small-dotted line in the loop denote the propagators 
 of the fermion and scalar components of $V_D$, respectively.
 The crosses denote the insertions of $F$ and $m$.  
 The propagators of the $SU(2)_U$ and $SU(2)_C$ gauginos 
 are denoted by the vertical solid lines with the wavy curve.}
\end{figure}
\newpage
%
%
\begin{picture}(200,200)
\put(100,0){\begin{minipage}[t]{10cm}
\unitlength=0.10mm
\begin{picture}(1000,1000)(-350,100)
\bezier{50}(-600,0)(-575,75)(-550,0)
\bezier{50}(-550,0)(-525,-75)(-500,0)
\bezier{50}(-500,0)(-475, 75)(-450,0)
\bezier{50}(-450,0)(-425,-75)(-400,0)
\bezier{50}( 450,0)( 425, 75)( 400,0)
\bezier{50}( 500,0)( 475,-75)( 450,0)
\bezier{50}( 550,0)( 525, 75)( 500,0)
\bezier{50}( 600,0)( 575,-75)( 550,0)
\put(-600,0){\line( 1,0){200}}
\put( 600,0){\line(-1,0){200}}
\bezier{50}(0, 150)( 75, 175)(0, 200)
\bezier{50}(0, 200)(-75, 225)(0, 250)
\bezier{50}(0, 250)( 75, 275)(0, 300)
\bezier{50}(0, 300)(-75, 325)(0, 350)
\bezier{50}(0, 350)( 75, 370)(0, 400)
\bezier{50}(0,-150)(-75,-175)(0,-200)
\bezier{50}(0,-200)( 75,-225)(0,-250)
\bezier{50}(0,-250)(-75,-275)(0,-300)
\bezier{50}(0,-300)( 75,-325)(0,-350)
\bezier{50}(0,-350)(-75,-370)(0,-400)
\put(0,-400){\line( 0, 1){250}}
\put(0, 400){\line( 0,-1){250}}
\put(-392,0){\line( 1, 1){400}}
\put(-407,0){\line( 1, 1){400}}
\put( 392,0){\line(-1,-1){400}}
\put( 407,0){\line(-1,-1){400}}
\put(-400,   0){\circle*{20}}
\put(-350, -50){\circle*{20}}
\put(-300,-100){\circle*{20}}
\put(-250,-150){\circle*{20}}
\put(-200,-200){\circle*{20}}
\put(-150,-250){\circle*{20}}
\put(-100,-300){\circle*{20}}
\put( -50,-350){\circle*{20}}
\put(   0,-400){\circle*{20}}
\put(400,  0){\circle*{20}}
\put(350, 50){\circle*{20}}
\put(300,100){\circle*{20}}
\put(250,150){\circle*{20}}
\put(200,200){\circle*{20}}
\put(150,250){\circle*{20}}
\put(100,300){\circle*{20}}
\put( 50,350){\circle*{20}}
\put(  0,400){\circle*{20}}
\bezier{200}(-150,0)(-140,-140)(0,-150)
\bezier{200}(-150,0)(-140, 140)(0, 150)
\put(150,  0){\circle*{10}}
\put(138, 57){\circle*{10}}
\put(106,106){\circle*{10}}
\put( 57,138){\circle*{10}}
\put(  0,150){\circle*{10}}
\put(138, -57){\circle*{10}}
\put(106,-106){\circle*{10}}
\put( 57,-138){\circle*{10}}
\put(  0,-150){\circle*{10}}
\put(-150,0){\line(1,1){50}}
\put(-150,0){\line(-1,1){50}}
\put(-150,0){\line(1,-1){50}}
\put(-150,0){\line(-1,-1){50}}
\put(150,0){\line( 1, 1){50}}
\put(150,0){\line(-1, 1){50}}
\put(150,0){\line( 1,-1){50}}
\put(150,0){\line(-1,-1){50}}
\put(-700,-150){\makebox(100,100){\Huge{$\lambda_{{}_N}$}}}
\put( 600,-150){\makebox(100,100){\Huge{$\lambda_{{}_N}$}}}
\put(250,200){\makebox(100,100){\Huge{$\tilde{P_i}$}}}
\put(-350,200){\makebox(100,100){\Huge{$\Psi_{P_i}$}}}
\put(-200,-650){\makebox(400,100){\Huge{Fig.1}}}
\end{picture}
\end{minipage}}
\end{picture}

\begin{references}
%
\bibitem{t'hooft}
 G. 't Hooft, 1979 Cargese Lectures, 
 published in 
 {\it Recent Developments In Gauge Theories, Proceedings}, 
 (NATO Advanced Study Institute New York, USA: Plenum, 1980).
%
\bibitem{rbs}
 L. Ibanez and G. G. Ross, 
 Phys.Lett. {\bf B110} (1982), 215.
 
 K. Inoue, A. Kakuto, H. Komatsu and S. Takeshita, 
 Prog. Theor. Phys, {\bf 68} (1982), 927.

 L. Alvarez-Gaume, M. Claudson and M. B. Wise, 
 Nucl. Phys. {\bf B207} (1982), 96. 
%
\bibitem{seiberg}
 N. Seiberg,
 Phys. Lett. {\bf B318} (1993), 469.

 N. Seiberg,
 Phys. Rev. {\bf D49} (1994), 6857.

 N. Intriligator, R. G. Leigh and N. Seiberg,
 Phys. Rev. {\bf D50} (1994), 1092.
%
\bibitem{dine}
 M. Dine and A. E. Nelson,
 Phys. Rev. {\bf D48} (1993), 1277.

 M. Dine, A. E. Nelson, and Y. Shirman,
 Phys. Rev. {\bf D51} (1995), 1362. 
 
 M. Dine, A. E. Nelson, Y. Nir, and Y. Shirman,
 Phys. Rev. {\bf D53} (1996), 2658.
%
\bibitem{kitazawa} 
 I. Affleck, M. Dine and N. Seiberg, 
 Phys. Lett. {\bf B137} (1984), 187.
 
 H. Murayama, 
 Phys. Lett. {\bf B355} (1995), 187.

 N. Kitazawa, 
 Nucl. Phys. {\bf B479} (1996), 336.
%
\bibitem{iy}
 K.-I. Izawa and T. Yanagida,
 Prog. Theor. Phys. {\bf 95} (1996), 949.
 
 K. Intriligator and S. Thomas,
 Nucl. Phys. {\bf B473} (1996), 121. 
%
\bibitem{ns}
 M. J. Strassler,
 Phys. Lett. {\bf B376} (1996), 119.

 A. E. Nelson and M. J. Strassler,
 Phys. Rev. {\bf D56} (1997), 4226.
%
\bibitem{okada}
 N. Kitazawa and N. Okada,
 Phys. Rev. {\bf D56} (1997), 2842.
%
\bibitem{ckn}
 A. G. Cohen, D. B. Kaplan and A. E. Nelson,
 Phys. Lett. {\bf B388} (1996), 588. 
%
\bibitem{o'r}
 L. O'Raifeartaigh, 
 Nucl. Phys. {\bf B96} (1975), 331.
%
\bibitem{hqu}
 M. Hqu, 
 Phys. Rev. {\bf D14} (1976), 3548.
%
\bibitem{randall}
 L. Randall, 
 Nucl. Phys. {\bf B495} (1997), 37.
%
\bibitem{mura}
 N. Arkani-Hamed and H. Murayama,
 Phys. Rev. {\bf D56} (1997), 6733. 
%
\bibitem{pdg}
 Review of Particle Properties, 
 Phys. Rev. {\bf D54} (1996), 1.
%
%
\end{references}
\end{document}